\documentclass[]{aastex631}
\usepackage[utf8]{inputenc}
\usepackage{amsmath}
\usepackage{graphicx}
\usepackage{subfigure}
\usepackage{amssymb}
\usepackage{amsthm}
\usepackage{subfigure}
\usepackage{xcolor}
\usepackage[figuresright]{rotating}
\usepackage{txfonts}
\usepackage{CJKutf8}
\usepackage{etoolbox}
\newcommand       \GBp          {G_{\rm BP}}
\newcommand       \GRp          {G_{\rm RP}}

\newcommand       \CBpRp        {(G_{\rm BP}-G_{\rm RP})_0}

\newcommand       \degree       {^\circ}

\shorttitle{Searching for M dwarfs in GLEAM-X DR1 catalog}
\shortauthors{Huang et al.}

\graphicspath{{./}{figures/}}
\begin{document}
\begin{CJK}{UTF8}{gbsn}
%%%%%%%%%%%%%标题

\title{Searching for radio late-type dwarf stars in the GLEAM-X DR1 catalog}

%%%%%%%%%作者
\correspondingauthor{Biwei Jiang}
\email{bjiang@bnu.edu.cn}

\author[0000-0002-4046-2344]{Qichen Huang(黄启宸)}
\affiliation{Institute for Frontiers in Astronomy and Astrophysics,
            Beijing Normal University,  Beijing 102206, China}
\affiliation{School of Physics and Astronomy, Beijing Normal University, Beijing 100875, Peopleʼs Republic of China}
\email{qchuang@mail.bnu.edu.cn}

\author[0000-0003-3168-2617]{Biwei Jiang(姜碧沩)}
\affiliation{Institute for Frontiers in Astronomy and Astrophysics,
            Beijing Normal University,  Beijing 102206, China}
\affiliation{School of Physics and Astronomy, Beijing Normal University, Beijing 100875, Peopleʼs Republic of China}

\author[0000-0002-3828-9183]{Zehao Zhang(张泽浩)}
\affiliation{Institute for Frontiers in Astronomy and Astrophysics,
            Beijing Normal University,  Beijing 102206, China}
\affiliation{School of Physics and Astronomy, Beijing Normal University, Beijing 100875, Peopleʼs Republic of China}

\author[0000-0002-3171-5469]{Albert Zijlstra}
\affiliation{Department of Physics and Astronomy, The University of Manchester, Manchester M13 9PL, UK}
%%\email{albert.zijlstra@manchester.ac.uk}

\begin{abstract}
%——————————————————————

We have developed a new method of multi-wavelength data combination for the search of late-type radio dwarfs, and have put it into practice using GLEAM-X DR1 data. The initial sample is selected by cross-matching the Gaia/DR3 objects with the probability of being a star no less than 99$\%$, and removing the extragalactic objects assigned by the SIMBAD database. The late-type dwarf stars are judged according to their location in the $(BP-RP)_0/M_{\rm G}$ color-magnitude diagram and in the $(J-H)_0/(K-W1)_0$ near-infrared color-color diagram. Furthermore, stellar activity is searched by ultraviolet excess in the GALEX/NUV band and the Rossby number in the TESS light curves. In total, 12 stars are found to be late-type dwarf stars associated with radio source, which is consisted of five stars with the UV excess and seven stars with the Rossby number less than 0.13. Three of these 12 stars are previously studied to be associated with radio objects. All these 12 stars are considered to be reliable counterparts of radio sources.

\end{abstract}

%——————————————————————


\section{Introduction} \label{sec:intro}

Detectable radio emission exist in stars of nearly all temperatures, masses, and evolutionary stages. The mechanisms for generating the radio emission are diverse, either thermal or non-thermal. Although radio emission usually accounts for a small portion of the overall stellar luminosity, studying radio wavelengths allows us to gain understanding about a wide range of stellar phenomena that cannot be investigated using any other methods \citep{gudel2002,matthews2018}.
For example, radio observations can be used to constrain stellar magnetic fields and plasma density in the magnetosphere, which is also an excellent measure of investigating the effect of stellar radio burst on planets. The radio emissions from exoplanets are currently the only way to measure the strength of their magnetic fields, which is an important factor of planetary habitability \citep{crosley2016,driscoll2015,airapetian2020}. 

As for late-type dwarfs, their radio emissions can be divided into two main components: weaker quiescent emission and brighter coherent radio bursts.  The quiescent emission is typically non-polarized or weakly polarized gyrosynchrotron emission generated by mildly relativistic electrons. By contrast, the coherent radio bursts are often highly circular polarized with brightness temperatures higher than $10^{12}$K, and they are produced by plasma emission or electron cyclotron maser (ECM) emission \citep{dulk1985,melrose1980,melrose2017}. The distinction between these two mechanisms lies in the former being generated by hotter solar-like stars, while the latter is produced by ultracool dwarfs or Jupiter-like objects, with the transition region between them lying in the M-dwarf sequence (approximately spectral types M3-M4). This transition may be related to the partly convective to fully convective transition inside stars \citep{yiu2024}. Further observational studies are needed to understand the relationship between these emission generation mechanisms and the variations in internal convection. Therefore, with the upcoming completion of SKA, it is important to establish a method that can search for and confirm radio stars on the large scale survey data.

Due to the weakness of radio emission from stars compared to other celestial objects, it has been challenging to capture radio emission from stars using previous radio telescopes under the sensitivity limitation. In 2017, The Murchison Widefield Array (MWA) was upgraded to 'Phase II,' which significantly improved its performance, extending the longest baseline to 5.5 km and increasing the angular resolution to 45 arcsecs. For the initial data release, the image RMS noise is 1.27 mJy/beam, reaching a level sufficient to detect late-type dwarfs' radio emission. The subsequent GaLactic and Extragalactic All-sky Murchison Widefield Array survey eXtended (GLEAM-X) survey project, targeting a wide range of observations in the declination southern of +30°, has created the possibility to search for radio stars within a large amount of radio data.

The key challenge of identifying stellar radio emission is chance coincidence with background radio galaxies. Direct position matches between the optical and radio surveys result in high chance coincidence probability. In the star catalogs obtained in the cross-matching of the LOFAR Two-meter Sky Survey (LoTSS) and Gaia DR2 within a range of 2.5$\arcsec$, only $\le10\%$ of the sources are identified as Galactic sample \citep{callingham2019}. To verify that the sources obtained from the cross-matching are stellar sources, it is necessary to pre-select the objects by using the physical properties of the radio and/or optical sources \citep{driessen2023}. Therefore, we need to filter and constrain the data in multiple ways to confirm the presence of radio stars within the dataset.

Currently, the main methods for identifying new radio stars make use of their characteristics, including circular polarization, variability, and proper motion. The circular polarization method searches radio stars through the circular polarization degree of radio emission, which is based on the characteristics that high circular polarization emission is essentially only present in radio stars. The synchrotron emission of the vast majority of active galactic nuclei (AGN) typically has less than $1\%$ circular polarization, greatly reducing the false positive rate during matching. The drawback is that many radio surveys currently lack the necessary circular polarization information, such as the GLEAM-X used in this work \citep{pritchard2021, callingham2021, toet2021}.  The variability method  relies on the intensity variation of radio emission caused by stellar flares to identify radio stars. This method not only excludes all stationary stars but also heavily depends on the sensitivity and resolution of radio telescopes. Such research is mainly conducted on individual radio stars \citep{driessen2020, driessen2022, andersson2022, fijma2024},  which is difficult to extend to a large number of sources. The proper motion method  identifies radio stars through the combined proper motion of stellar radio and optical emission. In comparison with the variability method, this is not limited to non-thermal emissions and flares.  But radio surveys with sufficient sensitivity, such as FIRST, have data for a time period not exceeding 25 years, which is relatively short for measuring the proper motion of most stars. Except for sources that are very close to us, the proper motion of stars is generally difficult to exceed the positional errors of radio surveys. This greatly limits the number of detectable sources \citep{driessen2023}.

% Due to the limitations such as resolution and field of view in radio, the quality and quantity of optical, infrared and ultraviolet observations are currently better than the radio data for stars. This work intends to take the advantage of large surveys in these bands to identify radio stars in the GLEAM-X/DR1 dataset. Similar to previous works, this work concentrates on the late-type dwarf stars that are generally active to emit radio emission at the frequency of the GLEAM-X survey. Simultaneously, their activities can be diagnosed in optical and ultraviolet bands. The procedure of this work includes to remove the non-stellar objects from the radio sources, to identify the late-type dwarf stars according to the color and brightness on the color-magnitude diagram, and to find the activity features in other bands. The data description is presented in Section \ref{sec:Data} followed by \ref{sec:Candidate search}. The results and the discussion are presented in Section \ref{sec:Results}.

Due to the limitations of resolution and field of view in radio, the quality and quantity of optical, infrared and ultraviolet observations are currently better than the radio data for stars. Therefore, using the advantages of other bands to search for radio stars is a better choice.
In this study, we develop a new method for searching for radio cool dwarfs based on the combination of multi-band data, which differs from other methods that heavily rely on radio data. We apply this method to the GLEAM-X DR1 data and conduct a search for radio cool dwarfs. The Gaia DR3 photometric data and 2MASS infrared data are used to identify stellar types, and the GALEX ultraviolet data and TESS light curve data are used to determine the activity level of the stars. These are detailed in Section \ref{sec:Identify stars} and Section \ref{sec:Activity} respectively. The data description is presented in Section \ref{sec:Data}, the results and the discussion are presented in Section \ref{sec:Results}.

%______________________________________________________%______________________________________________________%______________________________________________________%______________________________________________________%______________________________________________________
\section{Data} \label{sec:Data}

\subsection{The GLEAM-X/DR1 catalog}

As soon as the Murchison Widefield Array was upgraded to 'Phase II' in 2017, the GLEAM-X survey was initiated. Currently, the first batch of data is released, covering an area of 1,447 square degrees with 4h $\leq RA \leq$ 13h and $-32.7\degree \leq Dec \leq -20.7\degree$. This dataset contains a total of 78,967 sources, spanning 20 bands from 72 MHz to 231 MHz  with a resolution of 2$\arcmin$ – 45$\arcsec$ depending on the band. The peak and integrated flux data we used are wide-band data in the range of 170-231 MHz. The completeness of the catalog is $98\%$ at 50 mJy, with a reliability of $98.2\%$ at 5$\sigma$ and $99.7\%$ at 7$\sigma$. At 5.6 mJy, the completeness is still $50\%$. The astrometric offset in right ascension is $14\pm700$ mas, and  in declination is $21\pm687$ mas \citep{hurley2022}. It is in this catalog that we search for radio dwarf stars.

\subsection{Removal of non-stellar objects}

Previous studies show that majority of radio sources are non-stellar. The first step in the study is to remove the apparent non-stellar objects from the Gaia and SIMBAD database.

\begin{figure}
    \centering
    \includegraphics[scale=0.45]{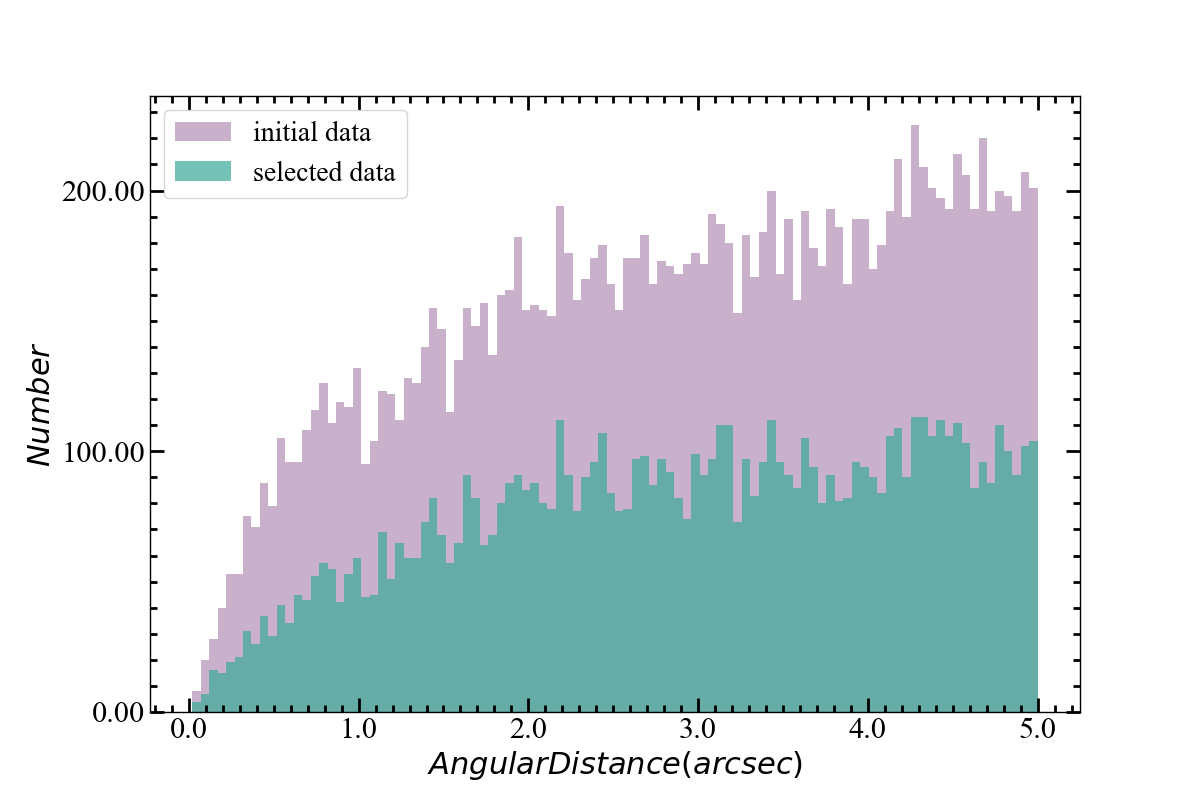}
    \caption{The result of the cross-match between Gleam-X and Gaia/DR3 within a 5 arcsecond radius shows that the initial data in purple which contains 15,365 sources and the selected data in green reduced to 7,654 sources after filtering with the Angular distance $< ((E\underline{~}RA \times \cos{\delta})^2+E\underline{~}Dec^2)^{1/2}$, .}
    \label{fig:ArcDis}
\end{figure}

Gaia provides the largest catalog of astrometric information including the distance and proper motion as well as optical photometry, which makes it a highly reliable tool to obtain stellar parameters and identify late-type dwarf stars \citep{creevey2022}. For the searching of radio stars, the GLEAM-X/DR1 catalog is firstly cross-matched with the Gaia/DR3 catalog. After removing the sources with positional errors greater than 45$\arcsec$, within a radius of 5$\arcsec$ which is the average position uncertainty of GLEAM, 15,365 sources are found in the Gaia/DR3 catalog. Since GLEAM-X provides the position uncertainty for individual sources as $E_{RA}$ and $E_{Dec}$ in degrees, these values have been converted to arcseconds. The angular distance in the cross-matching with Gaia, $A$, is further required to fulfill the condition
\begin{equation}
    \begin{aligned}
        A < \sqrt{(E_{RA} \times \cos{\delta})^2+E_{Dec}^2)}
    \end{aligned}
\end{equation}
to minimize the wrong matches, where $\delta$ is the declination. After applying this filtering criterion, there are still 7,654 sources, the distribution of the filtered sources is shown in Figure \ref{fig:ArcDis}. It should be noted that such 1-sigma constraint on the angular distance is relatively stringent so that some stars must have been removed during this process, which means that the resultant sample is not complete, and the distribution of GLEAM-X position error is neither Gaussian.

The sources that are matched with the Gaia/DR3 catalog are further cleaned by cross-matching with the SIMBAD database to remove the known non-stellar objects. The Gaia coordinates is used for its much more accurate position instead of the GLEAM-X. With a radius of 1$\arcsec$, a total of 864 matches are found in the SIMBAD database, among which 810 are classified as non-stellar such as AGN and galaxies. After removing these non-stellar objects, 6844 sources are left for further screening.

These 6844 sources are further cleaned by the parameter $classprob\_dsc\_combmod\_star(Pstar)$ of Gaia/DR3. For 219 millions targets, Gaia/DR3 uses Specmod method to obtain the probabilities of being a single star, a white dwarf, a binary star, a galaxy, or a quasar. These probabilities reflect a prior distribution of celestial bodies inside and outside the Milky Way and generate posterior class probabilities \citep{creevey2022}. The parameter $classprob\_dsc\_combmod\_star(Pstar)$  describes the target source probability of being a star, whose distribution actually peaks at the high and low end for stars and non-stars respectively. Only the objects with $Pstar \geq 0.99$ are selected as potentially stellar sources, while those which fall below this value are identified as non-stellar sources. The option of $Pstar\geq 0.90$ is checked that brings about only slight increase of number. With this choice, the sample is reduced to 5,939 sources.

%So, based on the target source probability of being a star :$classprob\_dsc\_combmod\_star(Pstar)$ provided by Gaia/DR3, these data are further cleaned by us. We used a filtering criterion of $Pstar\le0.99$ to remove all sources with stellar probabilities lower than this value, resulting in a final sample of 5,939 sources.
%
%To identify the real radio stars from these data, we conducted subsequent analyses combining the 2MASS, GALEX, and TESS datasets respectively.

%______________________________________________________%______________________________________________________%______________________________________________________%______________________________________________________
%______________________________________________________
\section{Identification of late-type dwarf stars} \label{sec:Identify stars}

%This work mainly focuses on late-type dwarf stars because compared to higher-temperature A-type stars and the majority of F-type stars, these stars are usually more active, have stronger magnetic field intensities, and are more likely to have strong radio emissions that we can detect. \citet{gudel2002} mentioned that flaring or quiescent radio emissions have been detected from G-K-type members of both the Pleiades and the Hyades.

\subsection{In the Gaia color-magnitude diagram} \label{subsec:Gaia}

The 5,939 sources selected above are assumed to be stars with $Pstar\ge 0.99$. In addition to late-type dwarf stars, this sample may include early-type dwarf stars and giant stars. In order to classify these sources, the color-magnitude diagram is used. Gaia provides the photometry in the $G$, $\GBp$ and $\GRp$ band for all the sources. Meanwhile, the parallax is available for 3820 sources, and only the sources with the parallax relative error $<20\%$ are kept for a reliable classification, which results in 1043 sources. Under this limitation on parallax, all extragalactic sources should be completely excluded. In addition, the extinction correction is necessary to correctly identify the stellar type.  Although Gaia/DR3 provides extinction information for some sources, they may suffer large uncertainty for the very red dwarf stars. Instead, the three-dimensional extinction map by \citet{green2019} is adopted by using the Gaia parallax to obtain the extinction towards these sources. With the value of $E(B-V)$, the extinction in the $G$ band and the color excess $E(\GBp-\GRp)$ are calculated with the conversion by \citet{wang2019}, i.e. $A_{\rm G}=2.49E(B-V)$ and $E(\GBp-\GRp)=1.3E(B-V)$. The distance is simply the inversion of the parallax.

\begin{figure}
    \centering
    \includegraphics[scale=0.45]{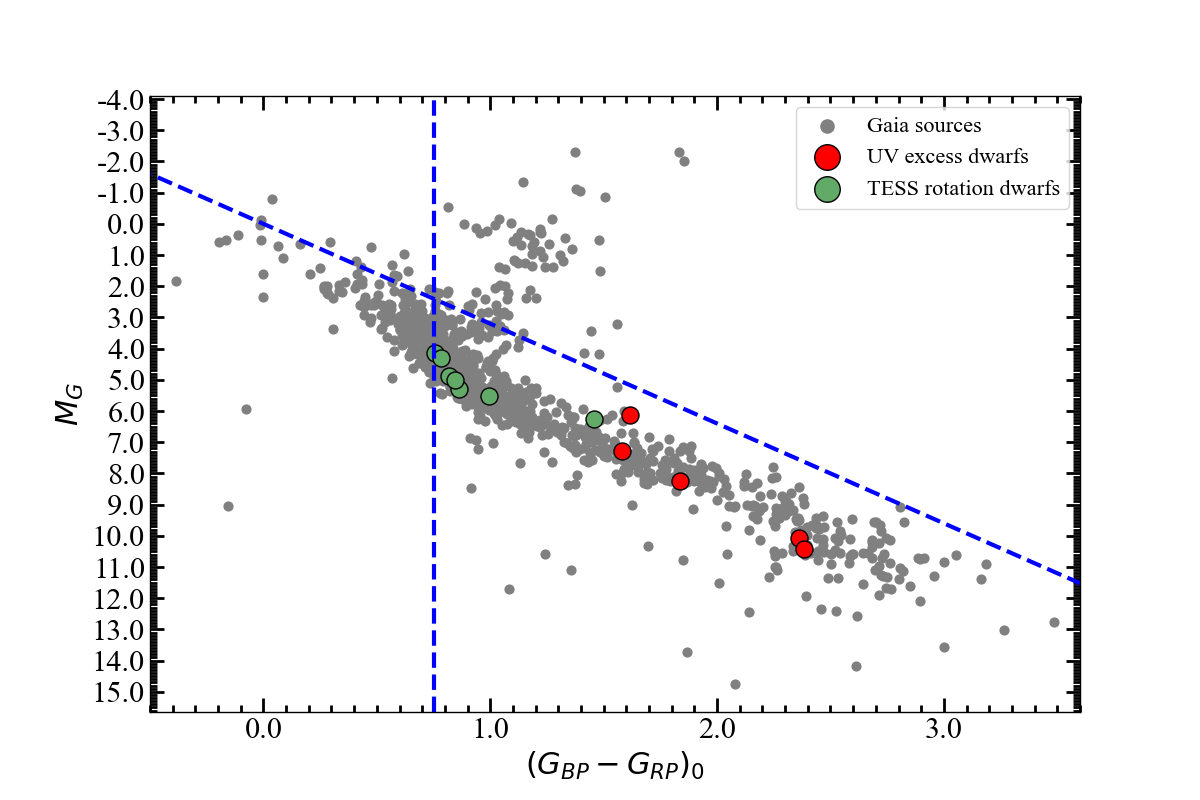}
    \caption{The Gaia color-magnitude diagram of the sample. The x-axis is $(\GBp-\GRp)_0$ color index after subtracting the interstellar extinction effects, and the y-axis is the absolute magnitude in the Gaia G-band. The color bar decodes the extinction in the G-band. The red and green dots are the sources which we identified as radio late-type dwarfs by UV excess and TESS rotation period, respectively.
    }
    \label{fig:Gaia CMD}
\end{figure}

The color-magnitude diagram of the 1043 stars is displayed in Figure \ref{fig:Gaia CMD}, which well characterizes the main sequence and the red giant branch. From the view of color, the intrinsic color index of $(\GBp-\GRp)_0$ equals to 0.75 at an effective temperature of 6000K for a G0-type star \citep{sun2023}. Thus $(\GBp-\GRp)_0 \geq 0.75$ is required to select late-type stars. The giant stars are excluded by the appearance in the CMD, i.e. the absolute magnitude in the $G$ band and intrinsic color index of $(\GBp-\GRp)_0$, i.e. $M_{\rm G} \leq 3.2 \times (\GBp-\GRp)_0$. The 753 stars in the lower right part of Figure \ref{fig:Gaia CMD} that satisfy both criteria are considered to be late-type dwarf stars.

\subsection{In the 2MASS/WISE color-color diagram}

It is apparent that more than half of the star candidates cannot be classified in the Gaia color-magnitude diagram due to the lack or inaccurate measurement of distance. As a supplement, the near-infrared diagram is introduced to exclude the other type stars. The $J-H/H-K$ diagram is already proved to be an effective tool to separate dwarf and giant stars. The very early work by \citet{bessell1988} found that the red dwarfs stars have significantly smaller $J-H$ than the red giants (including red supergiant, red giant branch and asymptotic giant branch stars) at given $H-K$ due the $H$ band sensitivity of surface gravity. This property was used by
\citet{ren2021} to distinguish the foreground dwarf stars and member red giant stars in the M31 and M33 field of view. Here,  the $K$ band is replaced by the $WISE/W1$ band because the $2MASS/K_{\rm S}$ band sensitivity is about 2 magnitude lower than $W1$. At the same time, the $H$ band sensitivity to stellar surface gravity is retained so that the giant stars can be recognized. \citet{pecaut2013} calculated the intrinsic color indexes of the O9-M9 spectral type dwarfs in various infrared bands including the 2MASS and WISE bands, which is helpful for our identification.

In brief, the 2MASS $J$ (1.25$\mu$m), $H$ (1.65$\mu$m) point source detection limit is better than 15.8 and 15.1 mag, respectively. The combined catalog contains a total of 470 million sources, covering $\rm 99.998\%$ of the sky \citep{skrutskie2006}. The Wide-field Infrared Survey Explorer (WISE) W1 centered at 3.4$\mu$m, with its 5$\sigma$ photometric sensitivity being 16.6 Vega mag in unconfused regions \citep{wright2010}.

\begin{figure}
    \centering
    \includegraphics[scale=0.45]{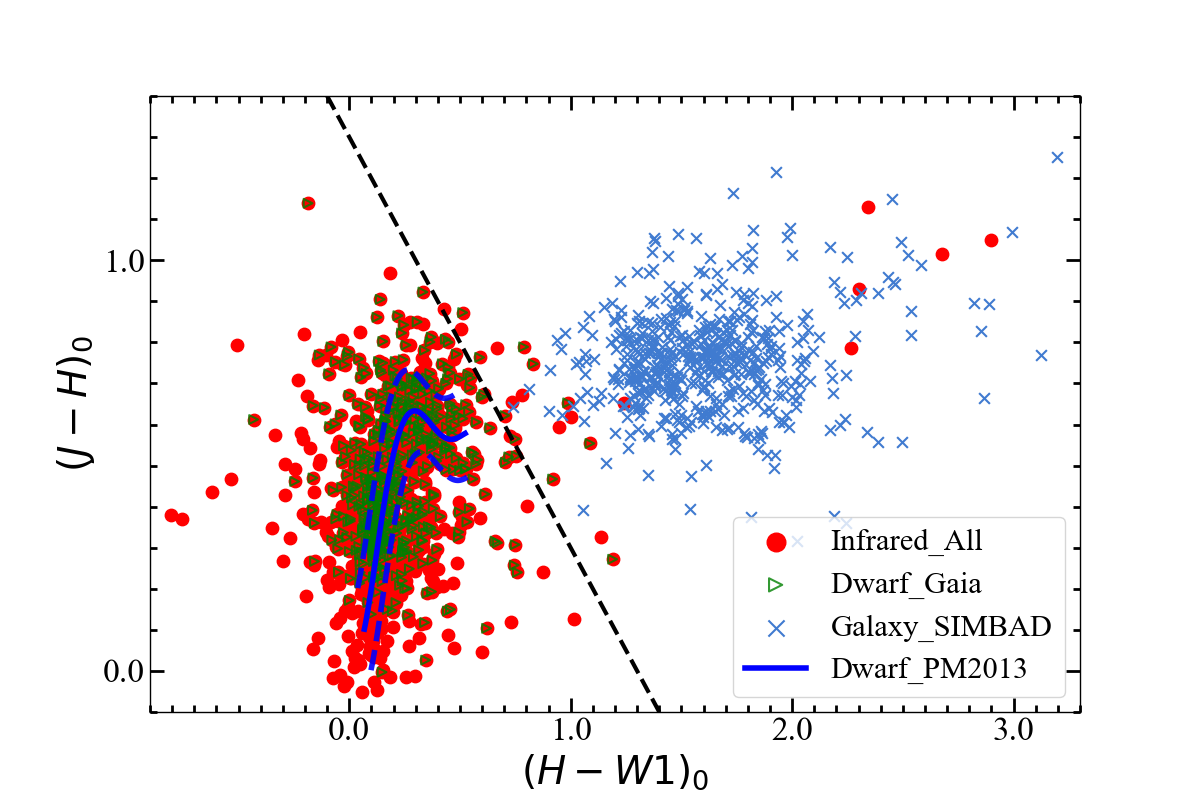}
   \caption{The near-infrared $(J-H)_0$/$(H-W1)_0$ diagram. The red dots represent the 988 sources selected by the quality screening of 2MASS and WISE observations, and the green dots represent the 541 dwarf stars selected by the Gaia CMD obtained in Section\ref{subsec:Gaia} with infrared data.  The blue crosses represent the 536 galaxies from the SIMBAD database with infrared data. The solid blue line represents the intrinsic color indices of F0-M5 spectral type dwarfs by \citet{pecaut2013}, where positions further to the right correspond to later spectral type stars, and the dashed lines above and below represent a 0.1-magnitude offset. The black dashed line represents the border for dwarf classification we set, i.e. $(J-H)_0 = (H-W1)_0 + 1.3$.
    }
    \label{fig:JHW1 CCD}
\end{figure}

Starting with all the 5939 sources obtained from the initial cleaning, the cross-match with the 2MASS and WISE catalog separately with a radius of 1$\arcsec$ yields 2207 and 2116 matches, respectively. To ensure the data quality, only the sources are retained from 2MASS with a quality of A or B in the J band and H band, and from WISE with a quality of A or B in the W1 band. With this criteria, the merged sample is consisted of 1135 sources. To calculate their extinction data, the Python Stellar Spectral Energy Distribution (PYSSED) was used \citep{mcdonald2024}. PYSSED routine combines photometry from disparate catalogues, to fit stellar luminosity, temperature, extinction and other stellar parameters. With PYSSED, the extinction is derived for 988 out of 1135 sources, for which the $(J-H)_0$ versus $(H-W1)_0$ diagram is displayed in Figure \ref{fig:JHW1 CCD}, as well as the dwarfs stars identified by the Gaia CMD, the galaxies identified by the SIMBAD database, and the position of dwarf stars calculated by \citet{pecaut2013} with a borderline of $\pm$0.1mag representative of 1$\sigma$ in the color index.

The dwarf stars are clearly separated from the other objects in Figure \ref{fig:JHW1 CCD}, which follow very well the color-color (blue lines) relation of \citet{pecaut2013}. The dwarf stars identified in the Gaia CMD (denoted by green dots) fall into this region as well. In contrast, the galaxies identified through the SIMBAD database are all much redder in $(H-W1)_0$.  Therefore, we manually draw a borderline, i.e. $ (J-H)_0 =  (H-W1)_0 + 1.3$, represented by a dotted line in Figure \ref{fig:JHW1 CCD},  where all the sources on the left of this borderline are dwarfs. Except galaxies, the sources in the right of the borderline may contain the dusty AGB stars \citep{ren2021} irrelevant to the topic of this work. This process identifies a total of 963 dwarfs. To make sure the dwarfs found are late-type stars, the luminosity provided by PYSSED is used to check for all the dwarfs. The luminosity criteria is derived from \citet{carroll2017}, the sources whose luminosity is higher than F8-type star (1.7 solar luminosity) are excluded. After this processing, 563 late-type dwarfs are identified.

%%%Unlike in the Gaia CMD criterion where the stars with temperatures higher than G0-type are excluded, these stars are not excluded in the near-infrared color-color diagram. This is partly because of the relatively large infrared photometric error in comparison with the Gaia/DR3. It is also partly because the much smaller difference between various types in the near-infrared approximating the Rayleigh-Jeans tail, so that a small interstellar extinction can shift the stars in possible type identification.

A total of 753 and 563 late-type dwarf stars are identified through Gaia and near-infrared photometry, respectively. Of these, there are 456 sources identified by both methods, 297 sources identified only by Gaia, and 107 sources identified only in near-infrared.
The 753 late-type dwarf stars identified from the Gaia CMD and the 563 late-type dwarf stars identified in the 2MASS $\&$ WISE CCD are combined. Since the large portion of them overlap, the combined set contains a total of 860 sources.

\subsubsection{Verification of the Gaia/$Pstar$ parameter}

\begin{figure}
    \centering
    \includegraphics[scale=0.5]{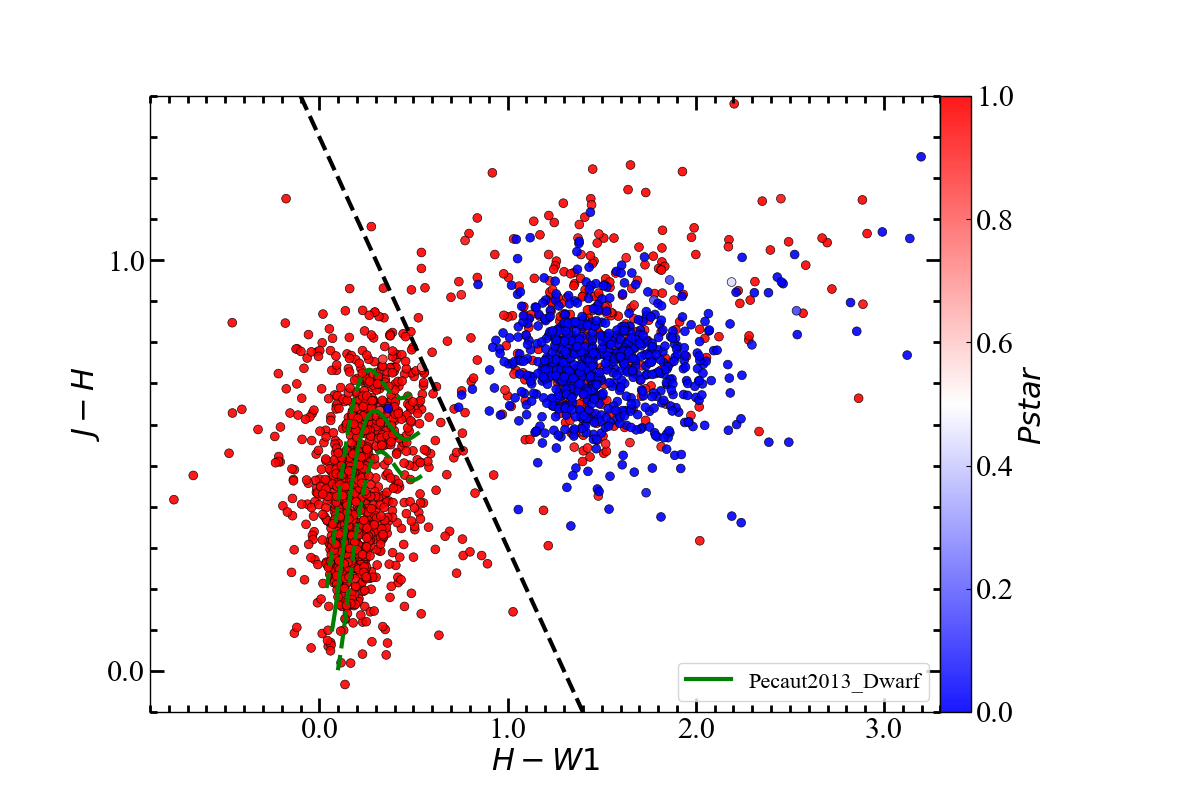}
    \caption{The J-H/H-W1 color-color diagram of 1989 sources displays the $Pstar$ value for each source. The black dashed line is consistent with the black dashed line in Figure \ref{fig:JHW1 CCD}, which represents the border for dwarf classification we set, i.e. $(J-H) = (H-W1) + 1.3$.}
    \label{fig:Pstar distribution}
\end{figure}

The stellar probability parameter $classprob\_dsc\_combmod\_star (Pstar)$ provided by Gaia which is used in the initial cleaning process is validated through the near-infrared data of 2MASS and WISE.
For the cross-matching catalog of GLEAM-X and Gaia, which contains 7654 sources, we perform a cross-matching of it with 2MASS and WISE without using $Pstar$ for selection. 3484 and 3536 sources are obtained from the two datasets, respectively. To maintain consistency, only sources with data quality of A or B in the J and H bands of 2MASS and data quality of A or B in the W1 band of WISE are retained. According to this criterion, the combined catalog of the two datasets contains a total of 1989 sources. For comparison, the combined catalog after $Pstar \geq 0.99$ screening has 1,174 sources remaining, i.e. there are  815 additional sources with $Pstar < 0.99$ in this catalog. This catalog, not filtered by $Pstar$, is also plotted in the J-H/H-W1 color-color diagram, as shown in Figure \ref{fig:Pstar distribution}.

It can be seen that 815 sources with $Pstar<0.99$ are mostly located at the right side of the black dashed line. Referring to the distribution of different sources in Figure \ref{fig:JHW1 CCD}, the vast majority of objects on this side are galaxies, which align with the selection criteria. On the left side of the dashed line, there is only one source with $Pstar<0.99$, indicating that this parameter as the initial screening condition for stars does not lead to the loss of late-type dwarfs which we aim to find.

To further confirm the types of sources distinguished based on $Pstar$, 815 sources with $Pstar<0.99$ are cross-matched with the SIMBAD database, obtaining results for 501 sources. Among them, 499 sources are classified as galactic objects (including quasars, galaxies, etc.), and 2 sources are classified as stellar objects, indicating that using $Pstar<0.99$ as an initial screening condition is reasonable.

\section{Search for stellar activity}
\label{sec:Activity}

The location in the optical color-magnitude diagram and in the near-infrared color-color diagram is a clear indication that the objects are stars. 
Late-type dwarf stars exhibit both quiescent emission and flaring emission. However, according to 
the observations at 4.8GHz by \citet{burgasser2005}, the quiescent emission intensities of two M-dwarf stars located at 10 to 20 pc were found to be 0.14 and 0.27 mJy respectively. \citet{pineda2023} observed the quiescent emission of a 3pc-distant M3.5 dwarf YZ Ceti at 2-4GHz to be 0.313 mJy. For our sources, the nearest distance is about 400 pc, which is much further than the aforementioned stars. Therefore, according to the sensitivity of GLEAM-X at 1 mJy/beam, it is very hard to detect such weak quiescent emission.
Therefore, late-type dwarf stars in this work is mostly detectable for their strong activities such as flares, and the active evidence is searched among these candidate radio stars. The search is performed in two ways, one is the ultraviolet excess, and the other is the optical light variation.

\subsection{The GALEX/NUV excess} \label{subsec:UV excess}

%%In the previous sections of Gaia/DR3 and the Infrared part, we have completed the selection of late-type dwarf stars. Therefore, our further goal is to identify sources with strong activity among these late-type dwarf stars as candidates for radio stars.

The "UV excess" refers to the stellar luminosity in the UV band exceeding its expected level while remaining normal in other bands, which can be manifested by a smaller color index UV-$[\lambda]$. For late-type stars in this work, it refers to UV emission above the photospheric contribution.
The reason for this excess can be attributed to the non-uniform magnetic fields in the deep convective zone of low-mass stars. These magnetic fields rise above the stellar surface and discharge a large amount of energy in the chromosphere, resulting in steady or variable emission \citep{shkolnik2014,miles2017,loyd2018}. This UV excess effect allows active dwarf stars to be significantly distinguished from other sources in the ultraviolet wavebands.

The UV data are taken from the Galaxy Evolution Explorer (GALEX) survey. Its AIS (the All-sky Imaging survey) has typical depth of 20/21 mag (FUV/NUV, in the AB mag system). The GALEX DR6+7 AIS catalog contains a total of 82,992,086 sources and is used in this work \citep{martin2005}.

\begin{figure}
    \centering
    \includegraphics[scale=0.45]{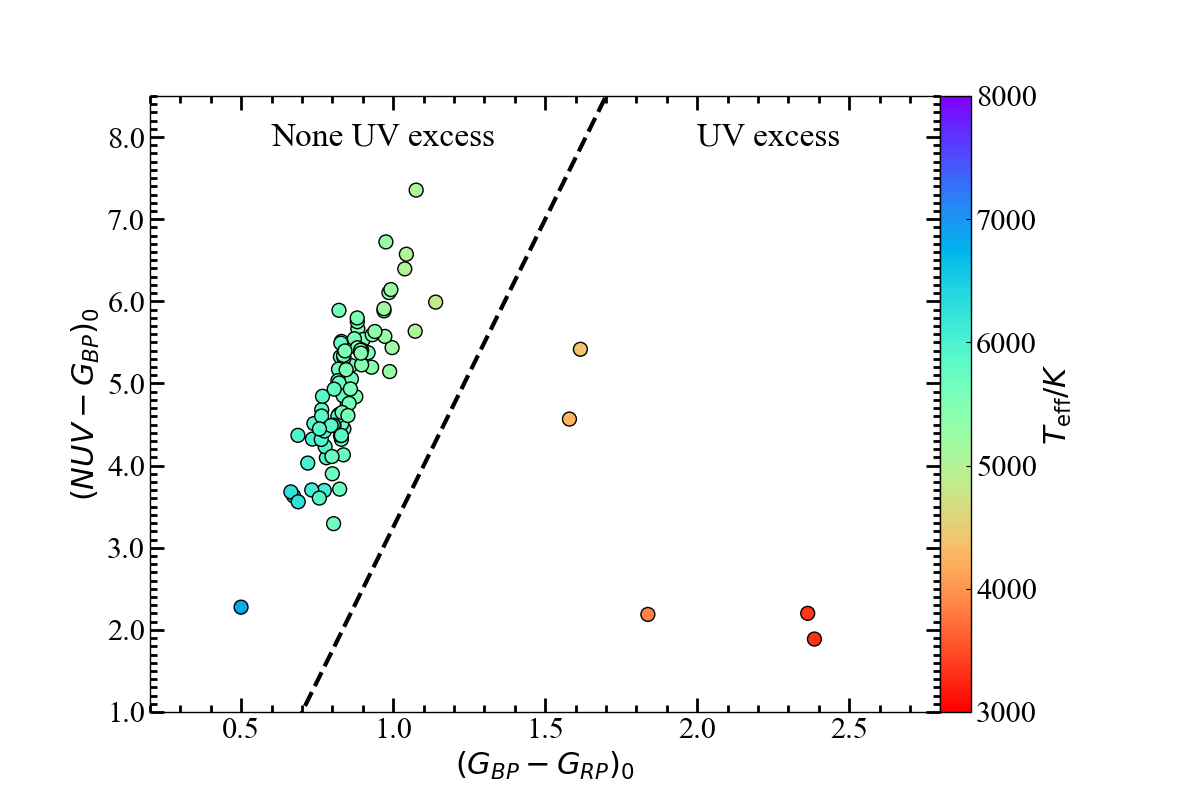}
    \caption{The  $(NUV-\GBp)_0$ vs. $(\GBp-\GRp)_0$ diagram for the 89 sources cross-matched with the GALEX/DR6+7 catalog. The color bar encodes the effective temperature. The black dashed line distinguishes five sources with ultraviolet excess on the right side.
    }
    \label{fig:UV excess}
\end{figure}

The 860 stars identified in the optical CMD and/or near-infrared CCD are investigated. With a 3$\arcsec$ radius in the Gaia coordinates, 89 sources are matched with the objects in the GALEX DR6+7 AIS catalog. The UV excess is to be identified from the excess in the NUV band because few sources are detected in the FUV band. For this purpose, the intrinsic color-color diagram of $(NUV-\GBp)_0$ vs. $(\GBp-\GRp)_0$ is built for these sources. The optical interstellar extinction is corrected with the $E(B-V)$ from PYSSED \citep{mcdonald2024} and the conversion to $E(\GBp-\GRp)$ from \citet{wang2019}. In addition, $E(NUV-\GBp)$ is calculated by using the conversion factor  $E(NUV-\GBp)/E(\GBp-\GRp)=3.25$ by \citet{sun2021}. The derived intrinsic color indexes yield the color-color diagram in Figure \ref{fig:UV excess}.

It is very clear in Figure \ref{fig:UV excess} that the sources in the color-color diagram are divided into two groups. One group includes most of the sources that follow an almost linear relation between the intrinsic color indexes $(NUV-\GBp)_0$ and $(\GBp-\GRp)_0$, which is expected from normal photospheric emission. On the other hand, the other group has five stars that show the NUV excess, an indicator of chromospheric activity. This group has systematically larger $(\GBp-\GRp)_0$, implying lower effective temperature. To quantify the effect of temperature, stellar effective temperatures are calculated by PYSSED, that determines the effective temperatures through SED-fitting and BT-SETTL AGSS2009 models \citep{allard2012}.

To double check, the color-spectral types relation of M-dwarfs by \citet{hejazi2020} is  used to infer their spectral subtype from the color index as following:
\begin{equation}
    \begin{aligned}
        \GBp-\GRp = 0.0183*{\rm SpT}^2 + 0.1620*{\rm SpT} + 1.7605
    \end{aligned}
\end{equation}
where $SpT$ is the spectral subtype number of M dwarf. The effective temperature at the designated spectral subtype is found in \citet{pecaut2013}. The results indicate that for an M0 type star, the temperature calculated by PYSSED is essentially accurate. For the two stars with the lowest temperatures in the calculation, the temperature calculated by PYSSED is about 3350K, while the spectral type calculated by \citet{hejazi2020} is around M3 corresponding to a temperature of 3400K, suggesting that the calculated temperatures of later type stars are also reliable.

With the effective temperature color-coded in Figure \ref{fig:UV excess}, it can be seen that the sources with temperatures higher than 5000K are concentrated on the left main branch, while the five sources with lower temperatures are distributed on the right. This coincides with the fact that the cooler dwarf stars are more active due to the increased opacity in the atmosphere,  which indicates the increased levels of magnetic activity to result in the increased brightness in the ultraviolet band, leading to smaller $(NUV-\GBp)_0$ under the same $(\GBp-\GRp)_0$ in comparison with normal stars.

\begin{figure}
    \centering
    \subfigure[]{\includegraphics[scale=0.39]{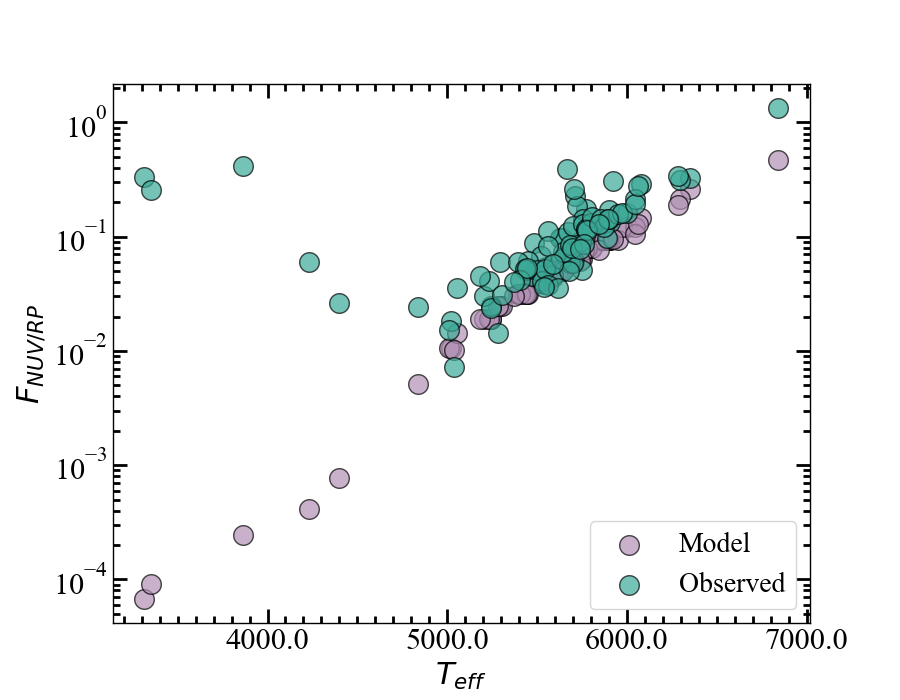}}
    %\hfill
    \subfigure[]{\includegraphics[scale=0.39]{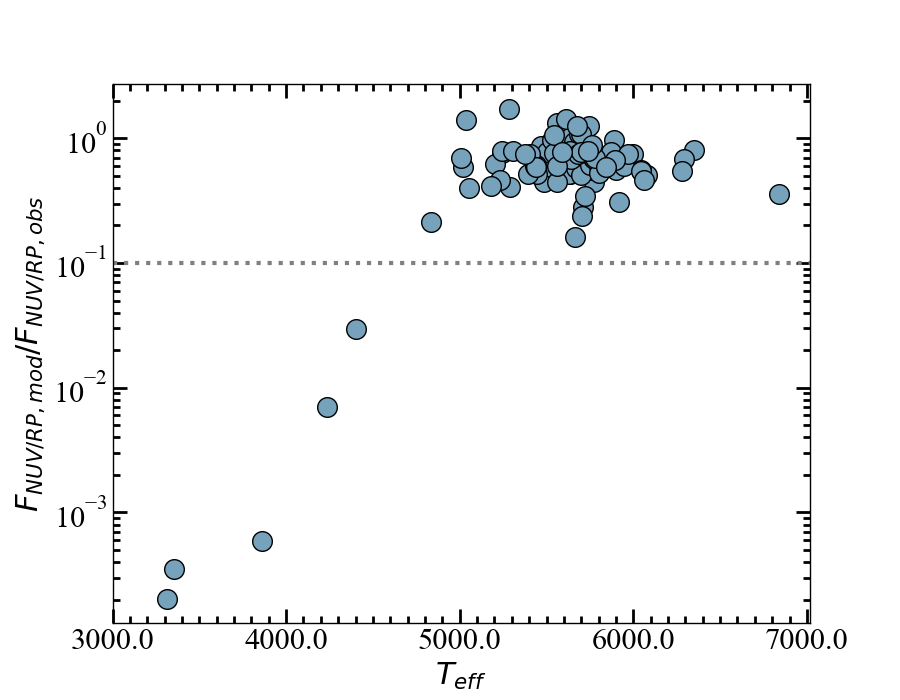}}
    \caption{The comparison between NUV fractional flux densities of model and observed.}
    \label{fig:NUV_RP compare}
\end{figure}

To further quantify the amount of UV excess, we refer to the method of \citet{shkolnik2014}. Firstly, we calculate each star's photometric flux by the PHOENIX model. As the model calculation does not include the emission from the chromosphere, transition region, and corona, the result is only the emission from the photosphere. Therefore, by subtracting the calculated flux from the observed, excess emission purely from atmospheric activity can be obtained.
Since the star is relatively distant and the calculation of the star's radius is somewhat inaccurate, we combine the data with the $\GRp$ band to analyze the NUV fractional flux densities, i.e., $F_{NUV}/F_{RP}$, thereby eliminating the impact of distance and star radius. The comparison between the model calculation and observation is plotted in Figure \ref{fig:NUV_RP compare}.

Assuming that the RP band is basically not affected by the chromosphere, five stars with temperatures below 4500K show a significant excess, as can be seen from Figure \ref{fig:NUV_RP compare}. For the three stars with temperatures below 4000K, the model-calculated fluxes (photospheric emission) are less than $0.1\%$ of the observed total emission. For stars with temperatures above 4500K, although most sources also exhibit total emission exceeding photospheric emission, none of them reaches a level exceeding ten times (dotted line in Figure \ref{fig:NUV_RP compare} (b).

These 5 stars are considered to be active UV excess stars and to be radio stars, and their information is recorded in Table \ref{tab:UV excess 5 stars}.

\subsection{The TESS light variation}

The activity of late-type dwarf stars can be distinguished not only through the ultraviolet excess, but also by their rotational period. \citet{wright2011} studied the relationship between the activity and rotation rates of 824 solar-type and late-type stars. They found that the activity indicated by X-ray-to-bolometric luminosity is correlated with stellar rotation rates and can be expressed by the Rossby number($R_O$). The $R_O$ is defined as $R_O \equiv R_P/\tau$, where $R_P$ is the rotation period and $\tau$ is the convective turnover time\footnote{$\tau$ is calculated through stellar mass derived from the mass-luminosity relationship.}. Their results showed that the relationship between stellar rotation rate and magnetic activity can be divided into two regimes: for slow rotators with $ R_O \geq 0.13 $, chromospheric activity increases sharply with decreasing rotation period, while for fast rotators with $ R_O \le 0.13 $, stellar activity becomes saturated and no longer changes with rotation rate. This result was confirmed in \citet{fang2018} and \citet{anthony2022}. 
In addition, \citet{fang2018} found that the Rossby number is a better factor than effective temperature to indicate the stellar activity. 
Therefore, for different late-type dwarf stars, determining activity through the $R_O$ calculated by period is feasible. Moreover, activity can serve as a criterion for determining flare rates. 
According to \citet{hawley2014}, although inactive stars can also exhibit flares, active stars tend to have higher flare frequencies and flare energies.
\citet{yiu2024} studied the impact of stellar flare occurrence on radio detectability using data from VLASS, LoTSS, GCNS, and TESS surveys.
They found that, the cumulative distribution function (CDF) curve of TESS flares sources is consistent with the CDF curve of VLASS/LoTSS detected sources at a confidence level of $> 95\%$ for stars above 3500K, although there is a significant inconsistency between the CDF of TESS flares sources and radio detected sources for samples below 3500K.
This implies that there is a significant correlation between TESS flare occurrence and radio detectability, for stars above 3500K.

In summary, for fast-rotating stars with temperatures above 3500K and $R_O < 0.13$ , as they exhibit stronger chromospheric activity, they are more likely to produce flares and consequently have a higher probability of being detected in radio surveys. The purpose in searching TESS data is to find stars that meet these two criteria. Using the rotation period data, we can assess the activity of the stars.

To derive the rotational periods of these stars, the Transiting Exoplanet Survey Satellite (TESS) photometric data are chosen for its frequent cadence that it conducts continuous photometric measurements for 27.4 days per cycle on a region of $24\arcdeg*96\arcdeg$. Currently it has covered most of the sky. During observations, it carries monitoring observations of all sources in the field at a cadence of 30-minute, and for some pre-selected sources, the cadence rises to 2-minute. At present, TESS has prepared the TESS Input Catalog (TIC) of over 1 billion objects \citep{stassun2019}.
The light curve data we used comes from MIT's Quick-Look Pipeline (QLP) \citep{huang2020tess}. It light curve data cover the full two-year TESS Primary Mission and include ∼ 14,770,000 and ∼ 9,600,000 individual light curve segments in the Southern and Northern ecliptic hemispheres, respectively.

The rotational period of stars can be derived through the TESS light curve because some late-type dwarf stars have surface spots that rotate with the star and bring about the periodic light variation.

As stated above, 860 sources are left after the selection by the Gaia CMD, near-infrared CCD, the relative error of parallax and stellar luminosity. The cross-match with the TESS Input Catalog v8.2 and QLP database yields 524 photometric light curves by using the Gaia coordinates and a 1$\arcsec$ radius.  To search for the sources with periodic light variation, the widely used $lightkurve$ package and $LombScargle$ in $astropy$ are taken to draw the periodograms and light curves for all sources. Assuming that the noise in the TESS data is white noise, all photometric data undergo three rounds of screening based on the standard normalized power values in the periodogram, signal-to-noise ratio, and FAP (false alarm probability). Firstly, the sources with a maximum normalized power value (ranging from 0 to 1) below 0.1 are all excluded, imposing strict restriction on the significance of periods in the light curve.  Secondly, the sources are filtered by a signal-to-noise ratio greater than 10, i.e. to exclude the sources with the strongest signal in the periodogram less than 10 times the average noise level (background noise). Subsequently, the sources are excluded if their maximum power value less than $FAP = 0.1\%$.  This criterion can be understood as assuming the absence of periodic signals in the data, there is only a $0.1\%$ chance of observing a signal higher than the FAP. Finally, all light curves are phased-folded and visually inspected to ensure that the signals are not due to instrumental or observational systematics or certain false positive signals.

\begin{figure}
    \centering
    \includegraphics[width=14cm]{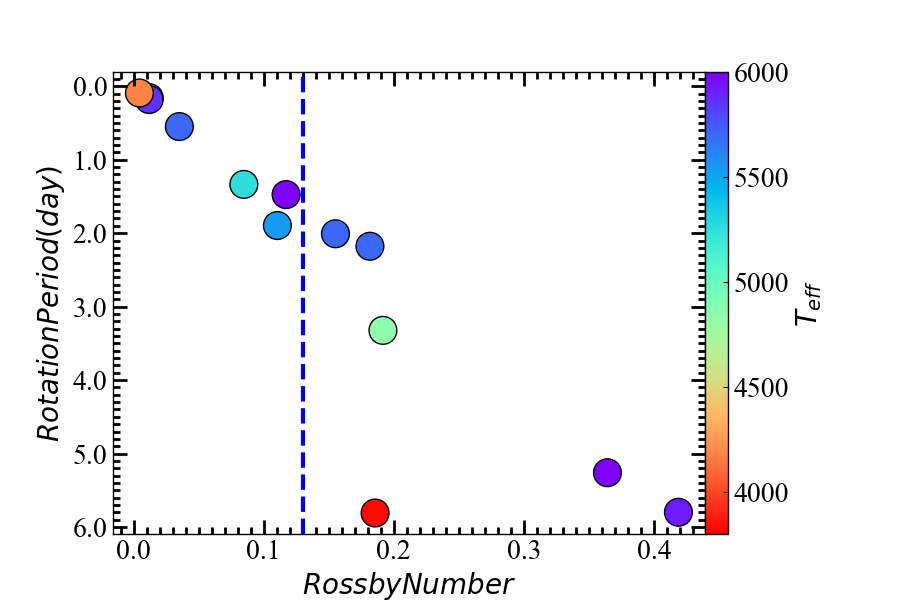}
    \caption{The relationship between Rossby number and rotation period for the 13 periodic variable late-type dwarf stars. The blue dashed line represents the boundary at $R_o = 0.13$, where stars to the left side of the dashed line can be identified as sources with stronger chromospheric activity. The colors indicate the effective temperature of the stars, with all temperatures being above 3500K, allowing for activity assessment based on the Rossby number.}
    \label{fig:Rossby period}
\end{figure}

\begin{figure}
    \centering
    \includegraphics[scale=0.35]{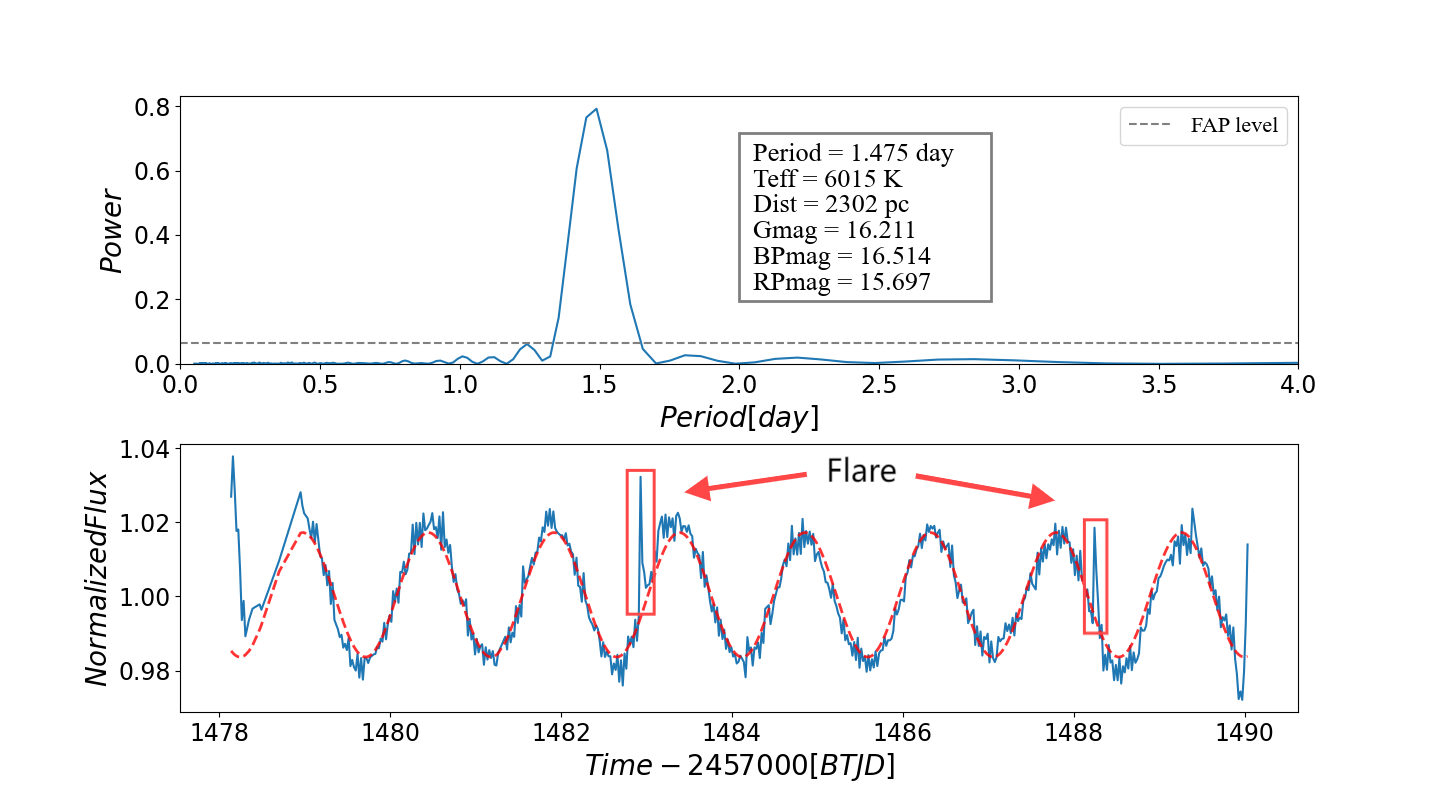}
    \caption{The light curve of TIC 124984182 within about 12 days, which shows obvious periodic light variation and some prominent flares, with a rotation period of 1.475 days and an effective temperature of 5795K corresponding to a G2-type star.
    }
    \label{fig:TIC 124984182}
\end{figure}

After all these verification, a total of 24 sources are found to show periodic variation. 
Among them, 11 sources are removed further because they are not late-type dwarf stars with the calculated effective temperatures higher than 6000K and $(J-H)_0 \le 0.258$, i.e. the G0 limit according to \citet{pecaut2013}.
The effective temperatures of the remaining 13 sources are all above 3500K, so they are screened based on the Rossby Number, as shown in Figure \ref{fig:Rossby period}. Among them, 6 sources with $R_o > 0.13$ are excluded. As a result, 7 sources are confirmed as candidate radio stars.
Among these 7 stars,  clear flares are detected in one source, TIC 124984182, and its light curve is displayed in Figure \ref{fig:TIC 124984182}. It can be seen that two prominent flares are present in two periods and some small flares in other periods. The effective temperature coincides with a G2-type star.

\begin{figure}
    \centering
    \includegraphics[scale=0.52]{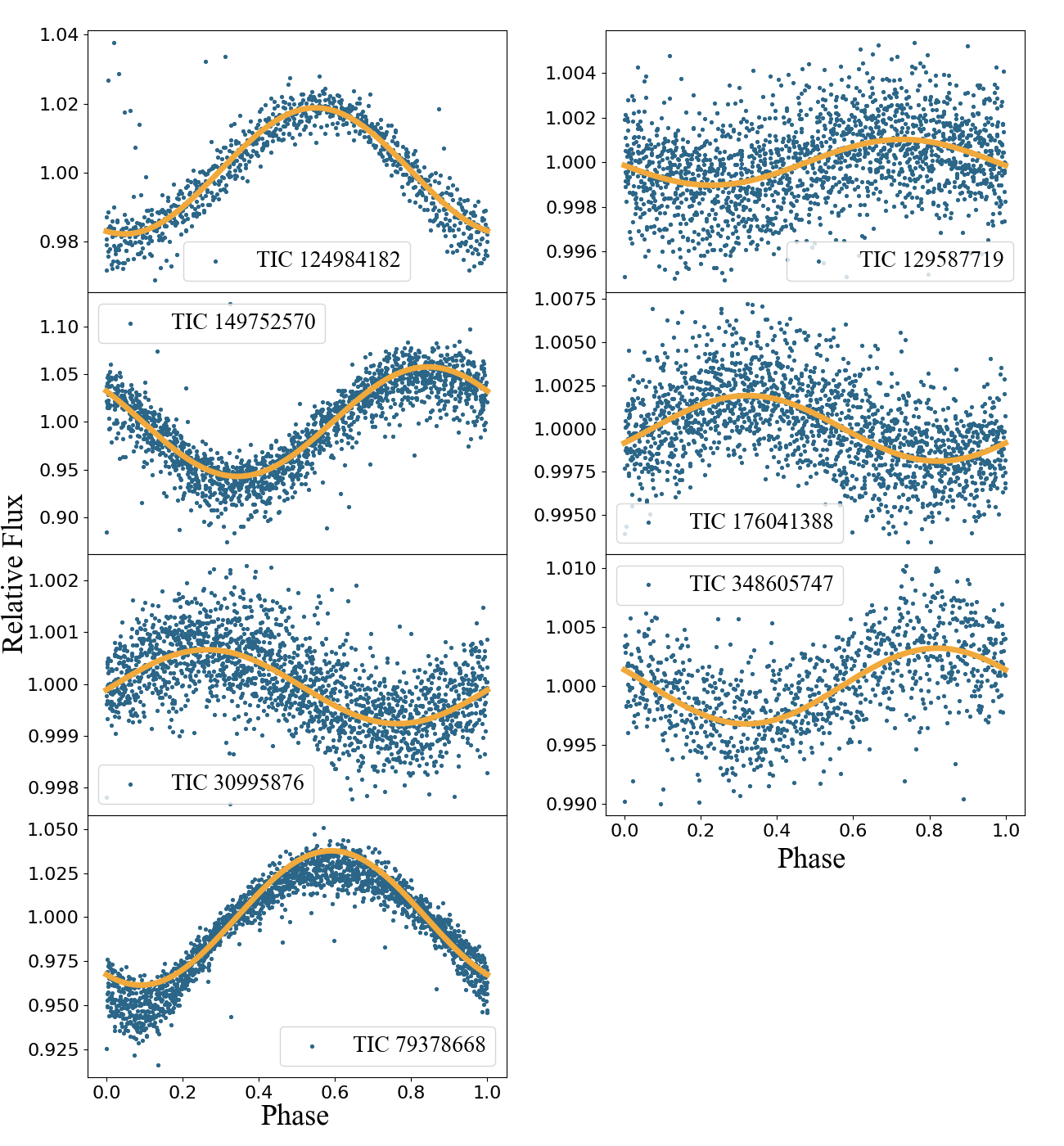}
    \caption{The phase-folded light curves of the 7 stars. The points deviating more than 3 sigma  are removed to eliminate some unreasonable data caused by the section switching of TESS.
    }
    \label{fig:Phase Fold combine}
\end{figure}

The light curves of all the 7 candidate radio stars are displayed through the phase-folded light curves in Figure \ref{fig:Phase Fold combine}, with all stellar parameters recorded in Table \ref{tab:TESS 7 stars}.

%______________________________________________________%______________________________________________________%______________________________________________________%______________________________________________________
%______________________________________________________

\section{Result \& Discussion} \label{sec:Results}

The GLEAM-X/DR1 objects are cross-matched with the Gaia sources with $Pstar\geq 0.99$. The late analysis of their Gaia color-magnitude diagram and the 2MASS/WISE near-infrared color-color diagram finds 860 late-type stars that may be the counterparts of the GLEAM-X radio sources. Subsequently, 12 highly active sources among late-type dwarf stars are selected through the UV excess and/or short rotational period  by using the GALEX NUV photometric data and TESS light curves. These sources are identified as candidate radio stars, with 5 sources selected through UV excess, and 7 sources through short rotational period. Their radio flux, magnitudes, stellar parameters, and other parameters are recorded in Table \ref{tab:UV excess 5 stars} and Table \ref{tab:TESS 7 stars}, respectively.

\begin{figure}
    \centering
    \includegraphics[width=14cm]{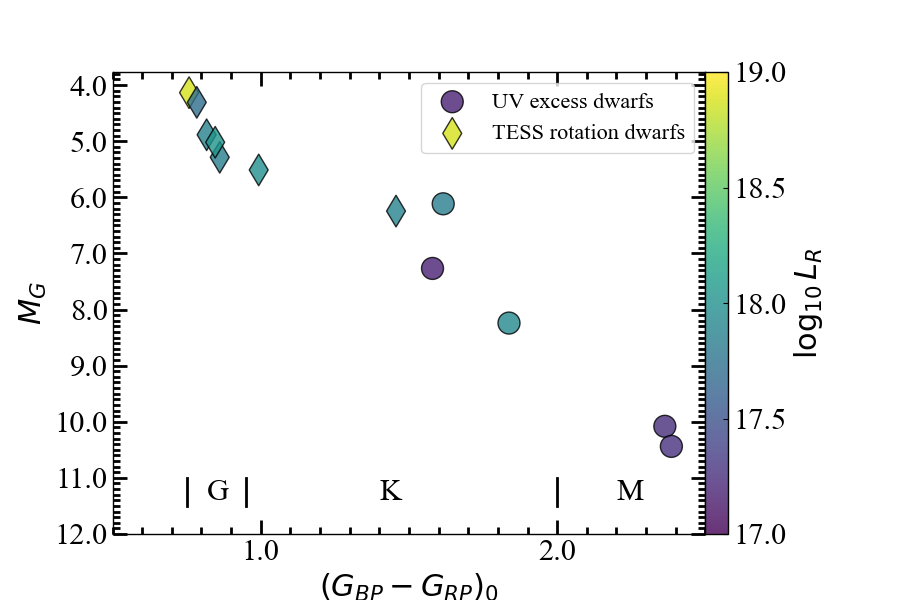}
    \caption{The distribution of all 12 identified radio stars on color-magnitude diagram, with colors indicating the radio luminosity calculated through distance from PYSSED and peak flux from GLEAM-X}
    \label{fig:Radio Luminosity}
\end{figure}

In Figure \ref{fig:Radio Luminosity}, the radio luminosity of each source is plotted according to the peak flux from GLEAM-X DR1, with the overall luminosity levels ranging from $10^{17}$ to $10^{19}$ (in units of $\rm erg \cdot s^{-1} \cdot Hz^{-1}$). When compared with the data in Figure 2 of \citet{pritchard2021}, it is more common for stars on the main sequence to have luminosities below $10^{18}$. Sources with luminosity above $10^{18}$ may represent special cases.

\subsection{Sources in the MORX v2}

The Millions of Optical - Radio/X-ray Associations (MORX) V2 is an update to the MORX catalogue \citep{flesch2016}. It is a catalog of radio/X-ray associations onto optical objects, using all the largest radio and X-ray source surveys available to 30 June 2023, such as, VLASS, LoTSS, RACS, FIRST, NVSS, and SUMSS radio surveys and having a total of 3,115,575 objects.

Three of the 12 stars are found as radio-optical counterparts in MORX v2 \citep{flesch2023}, specifically TIC 664536, TIC 6012946 and TIC 124984182, which supports our identification as radio stars. Their information, along with associated radio data, is presented in Table \ref{tab:MORX 3 stars}. This is the first time that the low-frequency data from the GLEAM survey is crossed with data from radio surveys for radio stars.
Another one of the 12 stars were studied by \citet{heinze2018}. They found that ATO J104.1998-22.4702 (TIC 79378668) is an eclipsing binary candidate.

No relevant research or records are found for the other 8 sources.

\subsection{Coincidence with AGN and quasar}

\citet{driessen2024} published The Sydney Radio Star Catalogue(SRSC), which contains 839 radio stars. To further investigate the coincidence between the identified radio stars and AGNs and quasars, we refer to their research and data to explore our results in-depth.
All 12 sources in the results are cross-matched with the Milliquas \citep{flesch2023Mill} and 6dFGSzDR3 \citep{jones2009} catalogues. The Milliquas contains approximately 1 million quasars and AGNs, and the 6dFGSzDR3 contains approximately 125,000 galaxies. The results showed that none of the 12 stars overlapped with any sources in these two catalogs within 3 arcmins. This supports that there is little chance coincidence between our sources and AGNs and quasars.

Regarding radio luminosity, the calculated radio luminosity is further compared with the SRSC result. Since the distances of 12 identified radio stars are generally greater than those in the SRSC catalog, we compare the results in Figure \ref{fig:Radio Luminosity} with the highest level data of different types of stars in Fig. 4 of \citet{driessen2024}. In the results of \citet{driessen2024}, a few G-type stars with $\rm BP-RP < 0.95$ and $\rm M_G = 4$ to $\rm 5$ reach a level of $10^{18} \rm erg \cdot s^{-1} \cdot Hz^{-1}$. For K-type stars with $\rm BP-RP = 0.95$ to $\rm 2.0$ and $\rm M_G = 6$ to $\rm 9$, about 5 sources reach a level of $10^{17}$ to $10^{18}$ $\rm erg \cdot s^{-1} \cdot Hz^{-1}$. For M-type stars with $\rm BP-RP = 2.0$ to $\rm 2.5$ and $\rm M_G = 9$ to $\rm 12$, fewer than 5 sources reach a level of $10^{17} \rm erg \cdot s^{-1} \cdot Hz^{-1}$. In summary, for GLEAM-X J062228.7-302648 (TIC 124984182) in our results, which has a radio luminosity of $10^{18.9} \rm erg \cdot s^{-1} \cdot Hz^{-1}$, its radio luminosity level is significantly higher than the same type stars. Since it exhibits obvious flares in the TESS light curve, we suspect that the higher radio luminosity is likely due to its strong activity. We hope to conduct further observational studies on this particular source. The vast majority of the remaining sources are below the level of $10^{18} \rm erg \cdot s^{-1} \cdot Hz^{-1}$, and the radio luminosity of two M-type dwarfs are $10^{17.2} \rm erg \cdot s^{-1} \cdot Hz^{-1}$, which is close to the extreme luminosity in \citet{driessen2024}.

%______________________________________________________%______________________________________________________%______________________________________________________%______________________________________________________
%______________________________________________________
\section{Summary}

We have developed a new method for searching for late-type radio stars by combining multi-band survey data. This method is based on the use of infrared and optical data to exclude extragalactic objects and non-late type dwarfs, and ultraviolet and time-domain data to select high-activity sources, thereby judging the possibility of the source as a radio star. The method is applied to the GLEAM-X DR1 radio survey data, combining it with the photometric data and time-domain data from Gaia/DR3, 2MASS, WISE, GALEX, TESS. As a result, 12 late-type stars are identified as radio counterparts, with distances ranging from 390pc to 2300pc. Among them, three sources are included in the MORX V2 database, and the remaining sources have not previously been identified as radio stars.

For future large-scale and high-precision radio surveys, our method will serve as an effective tool for screening the relatively distant radio stars. This will provide assistance for various studies like exoplanets and the radio emission mechanism of late-type dwarf stars.

\AtBeginEnvironment{acknowledgments}{\renewcommand\linenumbers[0]{}}

\section{Acknowledgments}

We are grateful to Drs. Tanda Li, Keyu Xing, Mengyao Xue, Prof. Tara Murphy, and the referee for their helpful discussion and suggestion. This work is supported by the NSFC project 12133002, National Key R\&D Program of China No. 2019YFA0405500, and CMS-CSST-2021-A09.  This work has made use of the data from GLEAM-X, 2MASS, GALEX, and WISE surveys.

\bibliographystyle{aasjournal}
\bibliography{main.bib}

\begin{sidewaystable}[]
 \caption{The radio stars identified by the UV excess}
    \centering
    \begin{tabular}{c c c c c c c c c c c c c c c c}
    \hline \hline
        GLEAM-X & RA & DEC & Plx & NUV & eNUV & $Gaia_{\rm G}$ & $A_{Gaia_{\rm G}}$ & $M_{Gaia_{\rm G}}$ & $\GBp$ & $\GRp$ & $\CBpRp$ & Int flux & Peak flux & $T_{\rm eff}^{Gaia}$ & $T_{\rm eff}^{this work}$ \\
         & ($^{\circ}$) & ($^{\circ}$) & kpc &  & & &  & & & & &  mJy & $\rm mJy/beam$ & K & K \\
        \hline

GALEX J043008.4-222256 & 67.5355 & -22.3818 & 0.90 & 22.21 & 0.39 & 16.35 & 0.123 & 5.99 & 16.66 & 15.00 & 1.59 & 8.4 & 6.5 &  & 4399 \\
GALEX J104816.3-304100 & 162.0682 & -30.6837 & 2.53 & 20.80 & 0.28 & 15.30 & 0.148 & 7.17 & 16.08 & 14.45 & 1.55 & 13.9 & 11.2 & 4464 & 4233 \\
GALEX J125627.0-230413 & 194.1123 & -23.0709 & 1.16 & 21.43 & 0.34 & 18.10 & 0.246 & 8.18 & 19.00 & 17.09 & 1.78 & 13.3 & 12.4 & 4147 & 3862 \\
GALEX J045307.7-292652 & 73.2827 & -29.4482 & 2.14 & 22.07 & 0.38 & 18.48 & 0.062 & 10.07 & 19.78 & 17.39 & 2.36 & 9.4 & 9.2 & 3435 & 3352 \\
GALEX J055744.4-263520 & 89.4343 & -26.5888 & 2.18 & 21.92 & 0.42 & 18.77 & 0.074 & 10.39 & 19.94 & 17.53 & 2.37 & 12.6 & 10.3 & 3479 & 3314 \\

    \hline
    \end{tabular}
    \label{tab:UV excess 5 stars}
\end{sidewaystable}

\begin{sidewaystable}[]
 \caption{The radio stars identified by the TESS light curve}
    \centering
    \begin{tabular}{c c c c c c c c c c c c c c c}
    \hline \hline
        GLEAM-X & RA & DEC & Plx & $Gaia_{\rm G}$ & $A_{Gaia_{\rm G}}$ & $M_{Gaia_{\rm G}}$ & $\GBp$ & $\GRp$ & $\CBpRp$ & Period & Int flux & Peak flux & $T_{\rm eff}^{Gaia}$ & $T_{\rm eff}^{this work}$ \\
         & ($^{\circ}$) & ($^{\circ}$) & kpc & &  & & & & & day &  mJy & $\rm mJy/beam$ & K & K \\
        \hline

GLEAM-X J062228.7-302648 & 95.6204 & -30.4457 & 0.40 & 16.21 & 0.049 & 4.20 & 16.51 & 15.70 & 0.79 & 1.475 & 13.1 & 13.5 & 5640 & 6015 \\
GLEAM-X J074600.3-312818 & 116.5017 & -31.4708 & 0.76 & 16.10 & 0.660 & 4.84 & 16.56 & 15.46 & 0.75 & 0.181 & 7.1 & 7.1 & 5516 & 5858 \\
GLEAM-X J065648.1-222809 & 104.1998 & -22.4703 & 1.16 & 14.12 & 0.111 & 4.33 & 14.45 & 13.60 & 0.79 & 0.155 & 6.5 & 7.6 & 5841 & 5720 \\
GLEAM-X J080527.2-232639 & 121.3640 & -23.4452 & 0.98 & 15.08 & 0.111 & 4.92 & 15.45 & 14.55 & 0.84 & 0.549 & 10.0 & 7.9 & 5670 & 5700 \\
GLEAM-X J052009.5-275259 & 80.0402 & -27.8824 & 1.00 & 15.40 & 0.172 & 5.22 & 15.77 & 14.86 & 0.82 & 1.896 & 7.6 & 7.8 & 5360 & 5540 \\
GLEAM-X J075658.7-290730 & 119.2449 & -29.1250 & 0.85 & 16.00 & 0.111 & 5.53 & 16.04 & 14.98 & 1.01 & 1.336 & 7.8 & 7.7 &   & 5255 \\
GLEAM-X J073449.1-235216 & 113.7036 & -23.8724 & 0.90 & 16.78 & 0.246 & 6.30 & 17.52 & 15.91 & 1.48 & 0.092 & 7.7 & 6.9 & 8714 & 4181 \\

    \hline
    \end{tabular}
    \label{tab:TESS 7 stars}
\end{sidewaystable}

\begin{sidewaystable}[]
 \caption{The radio stars matched in the MORX V2}
    \centering
    \begin{tabular}{c c c c c c c c c c c}
    \hline \hline
        GLEAM-X & RA & DEC & Int flux & Peak flux & \multicolumn{2}{|c|}{NVSS\footnote{1400 MHz}} & \multicolumn{2}{|c|}{RACS-mid\footnote{1367.5 MHz}} & \multicolumn{2}{|c|}{VLASS\footnote{3000 MHz}}\\

         & ($^{\circ}$) & ($^{\circ}$) &  mJy & $\rm mJy/beam$ & Name & Flux( mJy) & Name & Flux( mJy) & Name & Flux( mJy)  \\
        \hline

J125627.0-230418 & 194.1123 & -23.0709 & 13.3 & 12.4 &  J125626.9-230411 & 9.0 &  J125626.9-230412 & 8.9 &  J125627.08-230413.2 & 5.2 \\

J045307.7-292651 & 73.2827 & -29.4482 & 9.4 & 9.2 &  J045307.6-292650 & 22.2 &  J045307.6-292651 & 20.9 &  J045307.74-292650.5 & 13.5 \\

J062228.7-302648 & 95.6204 & -30.4457 & 13.1 & 13.5 &   &   &  J062228.7-302646 & 1.9 &   &   \\

    \hline
    \end{tabular}
    \label{tab:MORX 3 stars}
\end{sidewaystable}



\end{CJK}
\end{document}